\documentclass[useAMS,usenatbib]{mn2e} 
\usepackage{graphicx}

\title{Low-mass star formation in Lynds 1333}

\author[M. Kun et al.]{M. Kun$^{1}$\thanks{E-mail: kun@konkoly.hu}, S. Nikoli\'c$^{2}$,
 L. E. B. Johansson$^{3}$, 
 \newauthor
Z. Balog$^{4}$\thanks{On leave from University of Szeged, Dept. of Optics and Quantum Electronics,
D\'om t\'er 9, Szeged, H-6720 Hungary} and A. G\'asp\'ar$^{5}$  \\
$^{1}$Konkoly Observatory, H-1525 Budapest, P.O. Box 67, Hungary \\
$^{2}$Departamento de Astronom\'{\i}a Universidad de Chile, Casilla 36-D, Santiago, Chile \\ 
$^{3}$Onsala Space Observatory,  S-439 92 Onsala, Sweden \\
$^{4}$Steward Observatory, University of Arizona, 933 N. Cherry Av., Tucson AZ, USA 85721 \\
$^{5}$University of Szeged, D\'om t\'er 9, Szeged, H-6720 Hungary }

\begin{document} 

\date{Received  / Accepted }

\maketitle

\label{firstpage}

\begin{abstract}                                                                                
Medium-resolution optical spectroscopy of the candidate YSOs 
associated with the small, nearby molecular cloud Lynds~1333 revealed
four previously unknown classical T~Tauri stars, two of which are 
components of a visual double, and a Class~I source, 
\textit{IRAS}~02086+7600. The spectroscopic data, together with new
$V,R_{C},I_{C}$ photometric and 2MASS $J$, $H$, and $K_s$ data allowed us
to estimate the masses and ages of the new T~Tauri stars. We touch on
the possible scenario of star formation in the region. L\,1333 is one 
of the smallest and nearest known star forming clouds, therefore it may be 
a suitable target for studying in detail the small scale structure
of a star forming environment.
\end{abstract}

                                                                               
\section{Introduction}
\label{Sect_1}

Filamentary molecular clouds with embedded dense cores form a remarkable
subset of star forming clouds in our galactic environment 
\citep[e.g.][]{Onishi96,NC05,HTPG05}. Young stellar objects (YSOs) are 
associated with several cores along the filaments.
The formation scenario of the filaments and stars within them, however,
are not well understood. The filaments may be parts of shells, swept up by 
powerful stellar winds or supernovae \citep*[e.g.][]{KMT04}, or may result from  
fragmentation of sheet-like structures \citep{Hartmann02},
or may be shaped by large-scale flows like the galactic rotation \citep{Koda}.
Detailed studies of their density and velocity structures, as well as 
the properties of the YSOs born in them may help understand  their formation and
evolution.
 
Lynds~1333, a small dark cloud of opacity class 6 (Lynds 1962) in Cassiopeia,
at (l,b)=(128\fdg88,+13\fdg71) is part of a filamentary complex.   
According to the available observations L\,1333 is starless,
and thus has been included in several studies of starless cores 
\citep*[e.g.][]{LMT99,LMT01,Lee04}.  \citet*[][hereinafter referred to as Paper~I]{OKS}, 
studied first this cloud. They derived a distance of $180\pm20$\,pc 
from the Sun using Wolf diagram method. Their $^{13}$CO and C$^{18}$O observations
have shown L\,1333 to be part of a long, filamentary molecular structure, 
stretching from $l \sim 126\degr$ to $133\degr$ and from $b \sim +13\degr$
to  $+15\degr$, and referred to this molecular complex as {\em L\,1333 molecular cloud\/}. 
The angular extent of the molecular complex corresponds to a length of some 30\,pc
at a distance of 180\,pc. \citet{KMT04}  found that the L\,1333 complex is part 
of a giant far infrared loop GIRL~G\,126+10.
 
Recent star formation in the L\,1333 molecular cloud complex has been indicated by the 
presence of the \textit{IRAS}~source \textit{IRAS}~02086+7600, 
whose  \textit{IRAS} colour indices are indicative of a Class~I protostar, nevertheless
it coincides with a faint star in the  \textit{Digitized Sky Survey} image.
Due to its appearance as an optically visible star with large far-infrared 
excess several authors considered this object as a possible evolved star.
\citet*{FNP} included \textit{IRAS}~02086+7600 in a multiband photometric
survey for candidate post-AGB stars. They could not confirm the post-AGB nature 
of the star, and noted that it may be an ultracompact \mbox{H\,{\sc ii}} region, 
or a post-AGB star, or a YSO. \textit{IRAS}~02086+7600 appeared as a possible 
planetary nebula in the target lists of \citet{PM88} and \citet{VSP95}.

Based on its \textit{IRAS} colours, \citet{Slysh} included this object, as a candidate 
ultracompact \mbox{H\,{\sc ii}} region, in their search for OH maser emission.
They detected it as a thermal OH source at the velocity of 3.1km\,s$^{-1}$. 
The molecular maps presented in Paper~I revealed that this \textit{IRAS} source is 
projected on a dense C$^{18}$O core of a nearby molecular cloud whose radial velocity 
is +3.0\,km\,s$^{-1}$, same as that of the OH source, suggesting that  
\textit{IRAS}~02086+7600 most probably is a low-mass YSO. 
The C$^{18}$O spectrum observed at its position exhibited 
a wing-like feature, indicative of molecular outflow (Paper~I).
No known Herbig--Haro object is associated with this source. 

In addition to \textit{IRAS}~02086+7600, 18 H$\alpha$ emission 
stars have been detected in objective prism Schmidt plates in the region 
of L\,1333 (Paper~I). 
Three of these stars are associated with the \textit{IRAS} point sources
\textit{IRAS}~F02084+7605, 02103+7621, and 02368+7453. 

The aim of our present study is to establish an elementary  
data base on the star forming activity of the L\,1333 complex.
We observed the optical spectra of the candidate YSOs in order to establish their
pre-main-sequence nature and their spectral types. We also performed optical
photometry of the objects in order to determine their luminosities and positions in the HRD. 
We describe our observational data in Sect.~\ref{Sect_2}. Our results on the
properties of the observed stars, a short description of the large-scale 
environment of the cloud, and the possible star formation scenario are presented 
in Sect.~\ref{Sect_3}.  Sect.~\ref{Sect_4} gives a short summary.

\section{Observations and results}
\label{Sect_2}

\subsection{Spectroscopy}
\label{Sect_2.1}

All the PMS star candidate H$\alpha$ emission objects and \textit{IRAS} sources
listed in Paper~I were observed  on 4th January 2001, using the \textit{ALFOSC} 
spectrograph installed on the 2.5-m Nordic Optical Telescope in the Observatorio 
del Roque de los Muchachos in La Palma. The  
spectra were taken through grism~8, giving a dispersion of 1.5\,\AA/pixel
over the wavelength region 5800--8350\,\AA. Using a 1-arcsec slit
the spectral resolution was $\lambda / \Delta \lambda \approx 1000$ at 
$\lambda=6560$\,\AA. The exposure times of 900\,s for the H$\alpha$ 
emission stars resulted in $S / N \ga 100$. For the much fainter 
\textit{IRAS}~02086+7600 the exposure time was 2400\,s, 
resulting in  $S / N \approx 20$. 
Spectra of helium and neon lamps were observed before and 
after each stellar observation for wavelength calibration. We observed a 
series of spectroscopic standards for spectral classification purposes. 
\textit{IRAS}~02086+7600 was also observed on 13 September 2005, using the
CAFOS instrument on the 2.2-m telescope of Calar Alto Observatory. Using
the grism R-100, the observed part of the spectrum covered the wavelength 
interval 5800--9000\,\AA. The spectral resolution of CAFOS observation,
using a 1.5-arcsec slit, was $\lambda / \Delta \lambda \approx 1000$ at
$\lambda=8500$\,\AA. The exposure time  2400\,s resulted in  
$S / N \approx 7$ at 8500\,\AA. 
We reduced and analysed the spectra using standard {\sc iraf} routines. 

We confirmed the pre-main-sequence nature of three candidates listed in
Paper~I: OKS\,H$\alpha$~5, 6, and 16, all coinciding 
with \textit{IRAS} point sources and projected on the molecular clouds.
The other candidate H$\alpha$ objects listed in Paper~I proved to be field
stars without prominent H$\alpha$ emission and Li\,I absorption. 
We found by chance during the observations that a faint star some 1.8~arcsec 
south--southeast of OKS\,H$\alpha$~6, associated with \textit{IRAS}~02103+7621, 
was also a pre-main-sequence star. We refer to the two components as 
OKS\,H$\alpha$\,6\,N and OKS\,H$\alpha$\,6\,S, respectively. 

The wavelength range of \textit{ALFOSC} spectra was suitable for determining 
several flux ratios defined as tools for spectral classification 
by \citet{Kirk}, \citet{MK}, and  \citet*{PGZ}. We measured these spectral 
features on the spectra of our stars, and 
calibrated them  against the spectral type and luminosity class
by measuring them in a series of standard stars observed during the same run. 
The accuracy of the two-dimensional spectral classification,
estimated from the range of spectral types obtained from different
flux ratios, is $\pm1$ subclass 
\citet[for further details of spectral classification see][]{KPNJW}.

\begin{figure*}
\centering{
\includegraphics[width=16cm]{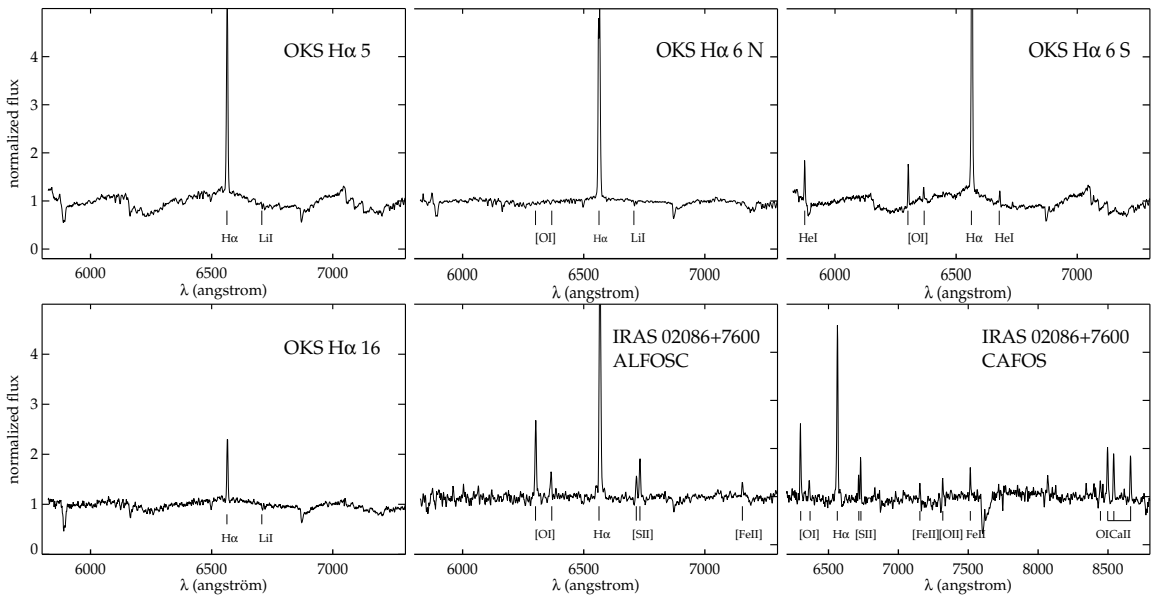}}
\caption{Optical spectra of the YSOs associated with the L\,1333 molecular complex.}
\label{Fig1}
\end{figure*}

Results of the spectroscopy are presented in 
Table~\ref{Tab1}. In addition to the derived spectral types we 
present the equivalent widths of the H$\alpha$ and \mbox{Li\,{\sc i}} lines in \AA, the 
10\%-width of the H$\alpha$ line in km\,s$^{-1}$, 
as well as list the additional emission lines observed in the
spectra. The uncertainties given in parentheses have been derived 
from the repeatability of the measurements. The real uncertainties of the 
\mbox{Li\,{\sc i}} equivalent widths may be higher due to the blending of the line with 
neighbouring absorption or emission features \mbox{(Ca\,{\sc i}}\,$\lambda$\,6718, 
\mbox{[S\,{\sc ii}]}\,$\lambda$\,6717). The spectra, normalized to the continua, 
are shown in Fig.~\ref{Fig1}.

\begin{table*}
\begin{minipage}{160mm}
\caption{Results of spectroscopy }
\label{Tab1}
{\footnotesize
\begin{tabular}{ll@{\hskip2mm}l@{\hskip1mm}r@{\hskip3mm}c@{\hskip3mm}c@{\hskip2mm}l}
\hline
Star & \textit{IRAS} name & Sp.T. & $EW$(H$\alpha$) & W(10\%)(H$\alpha$) & $EW$(Li\,{\sc i}) & Other emission lines \\
  &  &  & (\AA)~~~~  & (km\,s$^{-1}$) & (\AA) \\
\hline
 $\cdots$ & 02086+7600$^{1}$ & $<$ M0 & $-$44.8\,(0.5) & 570\,(10) &  $\cdots$ & \mbox{[O\,{\sc i}]}\,6300, 6363; 
\mbox{[S\,{\sc ii}]}\,6717, 6731; \mbox{[N\,{\sc ii}]}\,6548, \\
&&&&& &  6584; \mbox{[Fe\,{\sc ii}]}\,7155; \mbox{[Ca\,{\sc ii}]}\,7323; \mbox{[Ni\,{\sc ii}]}\,7378 \\
 $\cdots$ & 02086+7600$^{2}$ &  & $-$34.0\,(1.0) & 760\,(20) &  $\cdots$ & \mbox{[O\,{\sc i}]}\,6300, 6363; 
 \mbox{[S\,{\sc ii}]}\,6717, 6731; \\
&&&&& & \mbox{[Fe\,{\sc ii}]}\,7155; \mbox{[Fe\,{\sc ii}]}\,7319; \mbox{O\,{\sc i}}\,8446; \\
&&&&& &  \mbox{Ca\,{\sc ii}}\,8498, 8542, 8662 \\
OKS\,H$\alpha$\,5 &  F02084+7605 & M0.5IV & $-$25.1\,(0.5) & 610\,(12) & 0.47\,(0.02) & \\
OKS\,H$\alpha$\,6\,N & 02103+7621 & K7V & $-$51.5\,(0.5) & 710\,(10) & 0.40\,(0.02) & \\
OKS\,H$\alpha$\,6\,S & 02103+7621 & M2IV  &  $-$42.3\,(1.0) & 465\,(15) & 0.13\,(0.02) & 
\mbox{[O\,{\sc i}]\,6300, 6363}; \mbox{He\,{\sc i}}\,5873, 6678; \mbox{[S\,{\sc ii}]}\,6717, 6731 \\
OKS\,H$\alpha$\,16 & 02368+7453 & K7III & $-$8.0\,(0.2) & 520\,(15) & 0.71\,(0.02) \\
\hline
\end{tabular}}

\medskip 
$^{1}$ ALFOSC spectrum, 2001; $^{2}$ CAFOS spectrum, 2005
\end{minipage}
\end{table*}

Both Fig.~\ref{Fig1} and Table~\ref{Tab1} show that OKS\,H$\alpha$\,5,
OKS\,H$\alpha$\,6\,N, OKS\,H$\alpha$\,6\,S, and OKS\,H$\alpha$\,16 are classical T~Tauri stars 
(CTTS). Their spectral types are K or M, and their H$\alpha$ emission lines
fulfil the criteria established for the CTTSs by \citet{Martin} 
(i. e. $W$(H$\alpha$) exceeds the threshold value of 5\,\AA \ for K-type stars and 10\,\AA \ 
for M-type stars) and by \citet{WB03} (the width of the H$\alpha$ emission 
line 10\% above the continuum level is significantly larger than 300\,km\,s$^{-1}$). 
The spectral types were converted into effective temperatures 
$T_\rmn{eff}$ following \citet{KH95}  for luminosity class~V, and 
\citet{deJager87} for luminosity classes IV and III. The adopted $T_\rmn{eff}$ values are listed in
Table~\ref{Tab_result}.

\textit{IRAS}~02086+7600 displays an emission spectrum, containing strong 
H$\alpha$ and several forbidden lines, as well as the 
\mbox{Ca\,{\sc ii}} triplet in the CAFOS spectrum. 
The spectral resolution and $S/N$ of the spectra are insufficient for 
identifying absorption features, suitable for 
spectral classification. In particular, no TiO band, conspicuous in 
M-type spectra, can be seen, suggesting that its spectral type is probably earlier than
M0. The large number of forbidden lines resembles Class~I objects \citep{KBTB98,WH04},
thought to be either younger than, or identical with the youngest CTTSs. 
The high far-infrared excess of Class~I objects suggests
that the central star and its accretion disc are embedded in a dusty envelope.
Optical photons from such a source escape through the polar cavities of the
envelope, cleared by the protostellar wind. Several Class~I sources cannot be detected 
at optical wavelengths, and several others are  extended, suggesting that starlight, 
scattered from the circumstellar dust, contributes to their optical flux \citep[e.g.][]{Eisner}. 
The star-like appearance of \textit{IRAS}~02086+7600 suggests that one of its polar
cavities lies close to our line of sight. The extremely broad H$\alpha$ emission line
supports this assumption.

\subsection{\textbfit{V}, \textbfit{R}$_{\boldmath{C}}$, \textbfit{I}$_{\boldmath{C}}$ imaging and photometry}
\label{Sect_2.2}

Photometric observations of the young stars in L\,1333 in the $V$, $R_{C}$ 
and $I_{C}$ bands were undertaken
on 13 October 2001, 21 September 2003, and 11 December 2004 using the 1-m RCC-telescope of Konkoly
Observatory. In 2001 a Wright Instruments EEV CCD05-20 CCD camera was used, 
whose pixel size of 22.5\,$\mu$m corresponded to 0.35~arcsec on the sky. 
In 2003 and 2004 we used a Princeton Instruments VersArray:1300B
camera, that utilizes a back-illuminated, 1300$\times$1340 pixel Roper Scientific
CCD. The pixel size is 20\,$\mu$m, corresponding to 0.31~arcsec on the sky.
Integration times were between 180\,s and 600\,s. The open cluster NGC\,7790 
was observed each night several times, at various airmasses, 
for calibrating the photometry. We reduced the images in {\sevensize IRAF}. 
After bias subtraction and flatfield correction PSF-photometry 
was performed using the {\sevensize DAOPHOT} package. 
The transformation formulae between the instrumental and standard magnitudes
and colour indices as a function of the airmass were established each night by 
measuring the instrumental magnitudes of some 30 photometric standard 
stars published by \citet{Stetson} for NGC\,7790. 
OKS\,H$\alpha$\,6\,S was invisible in our $V$ images, thus only its
$R_{C}$ and $I_{C}$ could be determined. 




The results of the photometry for the three epochs are presented in Table~\ref{Tab2}. 
The photometric errors, given in parentheses, are quadratic sums of the formal errors 
of the instrumental magnitudes and those of the coefficients of the transformation 
equations. In some cases magnitudes measured at various epochs differ from 
each other by more than 0.1~mag. Comparison of the magnitudes of other stars 
within the field of the target objects has not shown such large 
discrepancies.  Therefore we conclude that part of these deviations is due to
the variability of the stars. In order to determine the interstellar extinction 
suffered by the stars and their luminosities we used the averages of the magnitudes
presented in Table~\ref{Tab2}.

\begin{table}
\caption{Results of optical photometry}
\label{Tab2}
{\footnotesize
\begin{tabular}{l@{\hskip1mm}c@{\hskip1mm}c@{\hskip1mm}c@{\hskip1mm}c@{\hskip1mm}c}
\hline
Star  & band & 13 Oct. 2001 &  21 Sept. 2003 & 11 Dec. 2004 \\
\hline
\textit{IRAS} 02086 & V & $\cdots$ & $\cdots$ & 19.65\,(0.07)\\ 
  & R$_\rmn{C}$ & 18.28\,(0.06) & 17.89\,(0.04) & 17.95\,(0.04) \\
&  I$_\rmn{C}$ & 16.65\,(0.03) & 16.36\,(0.04) & 16.55\,(0.04)  \\
\noalign{\smallskip}
OKS\,H$\alpha$\,5  & V & 15.54\,(0.05) & $\cdots$ & 15.38\,(0.04) \\ 
& R$_\rmn{C}$ & 14.12\,(0.04) & 14.02\,(0.03) & 14.03\,(0.03) \\ 
&  I$_\rmn{C}$ & 12.57\,(0.03) & 12.47\,(0.03) & 12.61\,(0.03)  \\
\noalign{\smallskip}
OKS\,H$\alpha$\,6\,N & V & 13.17\,(0.03) & 13.21\,(0.03) & $\cdots$ \\
  & R$_\rmn{C}$ & 12.59\,(0.03) & 12.50\,(0.03) & $\cdots$ \\
&  I$_\rmn{C}$ & 11.55\,(0.02) & 11.67\,(0.02) & $\cdots$ \\
\noalign{\smallskip}
OKS\,H$\alpha$\,6\,S & V & $\cdots$ & $\cdots$ & $\cdots$ \\
 & R$_\rmn{C}$ &  $\cdots$  & 15.20\,(0.10) & $\cdots$ \\
&  I$_\rmn{C}$ &   13.86\,(0.05) & 13.77\,(0.05) & $\cdots$  \\
\noalign{\smallskip}
OKS\,H$\alpha$\,16 & V &  $\cdots$ & 17.42\,(0.04) & 17.55\,(0.05) \\ 
 & R$_\rmn{C}$ & $ \cdots$ & 15.52\,(0.03) & 15.48\,(0.03) \\
&  I$_\rmn{C}$ &  $\cdots$ & 13.56\,(0.03)  & 13.55\,(0.03) \\
\hline
\end{tabular}}
\end{table}

Our images have revealed \textit{IRAS}~02086+7600 to be slightly extended.
Small reflection nebulae
can be seen next to both OKS\,H$\alpha$\,5 and \textit{IRAS}~02086+7600. 
We chose PSF-photometry in order to minimize the contribution of the extended 
emission to the resulting magnitudes listed in Table~\ref{Tab2}.

\section{Discussion}
\label{Sect_3}

\subsection{Interstellar extinction and SEDs}
\label{Sect_3.1}

We supplemented our observational data with the near infrared data of the 2MASS All 
Sky Catalog \citep{2MASS} and the far infrared data of the \textit{IRAS} 
PSC and FSC in order to characterize the circumstellar environments 
of our target objects. Other infrared, e.g. \textit{Spitzer} data are not available
for these objects. A single object is 
associated with OKS\,H$\alpha$\,6 in both catalogues. The fluxes of the counterparts,
2MASS\,J\,02152532+7635196  and \textit{IRAS}~02103+7621 are thus
combined from those of both components of the visual double.

We derived the interstellar extinction suffered by 
our stars based on the assumption that their total emission in the 
$I_{C}$ band originates from the photosphere \citep*[see e.g.][]{Meyer,Cieza05}, 
and the total $E_{R_{C}-I_{C}}$ colour excess results from interstellar 
reddening. The unreddened colour indices were adopted from \citet{KH95} for luminosity 
classes V and IV, and from \citet{Bessell} for OKS\,H$\alpha$\,16, whose spectral
features indicated a luminosity class~III.
The extinction  $A_{V}$ was derived from $E_{R_{C}-I_{C}}$ 
as $A_{V} = 4.76 \times E_{R_{C}-I_{C}}$ \citep{Cohen}.
For determining the extinction in the other photometric bands, we
used the relations $A_{R_{C}} = 0.78\,A_{V}$, 
$A_{I_{C}} = 0.59\,A_{V}$ for the optical \citep{Cohen}, and $A_{J} = 0.26\,A_{V}$,
$A_{H} = 0.15\,A_{V}$, $A_{K_{s}} = 0.10\,A_{V}$ for the near-infrared.
These latter values, slightly different from the standard \citet{RL} extinction law, are 
based on the relations presented for the 2MASS-bands by \citet{NC05}.



This method cannot be applied for \textit{IRAS}~02086+7600, whose
spectral type and thus photospheric colour indices are unknown. We derived 
the interstellar extinction suffered by this object by assuming 
that its unreddened position in $J-H$ vs. $H-K_{s}$ colour-colour diagram 
is on the {\em T~Tauri locus\/} defined by \citet{Meyer}, i.e. its colour
indices satisfy the relationship $(J-H)_0 = 0.58 (H-K)_0 + 0.52$.
We also determined  $A_{V}$ by this method for OKS\,H$\alpha$\,5 
and OKS\,H$\alpha$\,16 and found that the results were
compatible with those derived from $E_{R_{C}-I_{C}}$. 


The extinctions adopted are listed in column~2 of 
Table~\ref{Tab_result}. We note that the dereddened colour index $R_{C}-I_{C}$ of
\textit{IRAS}~02086+7600 is $(R_{C}-I_{C})_{0} = 0.73$, suggesting a $\sim$~K7-type star, 
in accordance with the absence of TiO bands in the spectrum.

\begin{figure}
\resizebox{\hsize}{!}{\includegraphics{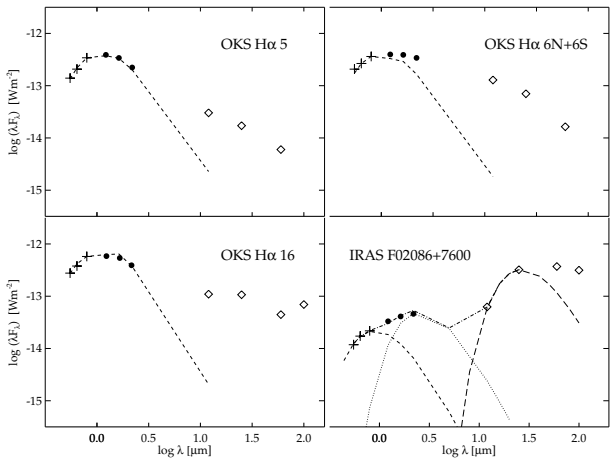}}
\caption{Spectral energy distributions, corrected for the interstellar extinction, of the 
YSOs associated with the L\,1333 molecular complex. Crosses result from our $V R_C I_C$
photometry, dots come from 2MASS data, and diamonds mark the \textit{IRAS} fluxes. 
Dashed lines show the contribution of the photoshere to the SEDs of the OKS\,H$\alpha$
stars, and a 4000\,K blackbody fitted to the optical fluxes of  \textit{IRAS}~02086+7600. 
In the plot of this latter object the dotted line shows a 1400\,K blackbody and the long-dash 
line is a 130\,K blackbody. The sum of these three components is drawn by dash-dotted line.}
\label{Fig_sed}
\end{figure}

We constructed the spectral energy distributions of the stars using their 
magnitudes corrected for the interstellar extinction. The fluxes corresponding to zero 
magnitude were obtained from
\citet{Glass} for the $V R_{C} I_{C}$ bands, and from the 2MASS All Sky
Data release web document\footnotemark[1]
\footnotetext[1]{ $\rmn{http://www.ipac.caltech.edu/2mass/releases/allsky/doc/}$
 \newline $\rmn{sec6\_4a.html}$}
for the $J H K_s$ bands. The resulting  dereddened SEDs are shown in Fig.~\ref{Fig_sed}.
In the plots of OKS\,H$\alpha$ stars, dashed lines show the SEDs of the photospheres, 
determined from the dereddened $I_{C}$ magnitudes and from the colour indices 
corresponding to the spectral types. 
The SEDs confirm the CTTS-nature of the OKS\,H$\alpha$ stars: their
SEDs display significant far-infrared excesses and negative slopes between 
2 and 25\,$\mu$m, characteristic of Class~II infrared sources \citep{Lada},
i.e. stars surrounded by dusty accretion disks. 
The plot of OKS\,H$\alpha$\,6 shows the sum of both components.

Contrary to the OKS\,H$\alpha$ stars, the slope of the SED of \textit{IRAS}~02086+7600 is  
$d \log (\lambda F_{\lambda}) / d \log \lambda = 0.80$ between 2 and 25\,$\mu$m,
characteristic of Class~I sources. The shape of the SED over the wavelength interval 
0.55--25\,$\mu$m can be well matched with the sum of three blackbodies, 
indicated in Fig.~\ref{Fig_sed}, and suggesting three dominant temperatures in 
the inner regions of the system. The hottest component, fitted to the optical part 
of the SED (dashed line), has $T \approx 4000\,K$ and corresponds to the 
photosphere of a K7-type central star. 
Ignorance of the contribution of veiling and scattered light to the  
optical fluxes make this temperature estimate somewhat uncertain. 
We rely on this value, in view of the observational results that the 
veiling is constant redward of $\sim$~5000\,\AA \ \citep[e.g.][]{BB90,WH04}, and 
assuming that we were able to exclude a considerable part of the scattered light by 
performing PSF-photometry. The uncertainty, estimated from the goodness of the 
fit, is $\pm200$\,K. The next component of the SED is a blackbody 
with $T = 1400\,K$, quite similar to the spectra of the 
NIR excesses of classical T~Tauri stars \citep{Muzerolle},  
which have been succesfully modelled as the `photosphere' of the inner rim of the
disc, emitting like a blackbody near the dust sublimation temperature.  
The third dominant temperature, suggested by the shape of the SED, is $\sim 130$\,K.
Beyond 25\,$\mu$m the SED turns flat, suggesting the peak position between 60 and 100\,$\mu$m.

\subsection{Positions of the pre-main-sequence stars in the HRD}
\label{Sect_3.2}

\begin{figure}
\resizebox{\hsize}{!}{\includegraphics{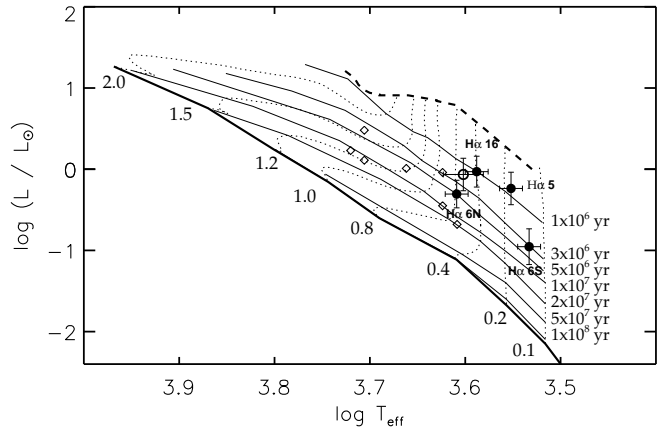}}
\caption{Positions of the young stars of the L\,1333 region in the HRD, assuming a distance
of 180\,pc. Black dots with error bars indicate the classical T~Tauri stars 
associated with L\,1333, and the open circle shows the estimated position of
\textit{IRAS}~02086+7600. Diamonds indicate the nearby weak-line T~Tauri stars 
identified by \citet{TNKF05}. Thin solid lines indicate the isochrones as labelled,
and dotted lines show the evolutionary tracks for the masses indicated at the 
lower end of the tracks according to \citet{PS99} model. The dashed line
corresponds to the birthline and thick solid line indicates the
zero age main sequence.}
\vskip -0.4cm
\label{Fig_hrd}
\end{figure}

Bolometric luminosities of the OKS\,H$\alpha$ stars were derived by applying the  
bolometric corrections $BC_{I_C}$, tabulated by \citet{Hartigan} to the 
dereddened $I_{C}$ magnitudes, and adopting a distance of 180\,pc. 
The distribution of the stars in the HRD is displayed in Fig.~\ref{Fig_hrd}.
Errors of $\log\,T_\rmn{eff}$ were derived from the accuracy of
$\pm\,1$ spectral subclass of the spectral classification.
The error of the luminosity comes from the quadratic sum of $\delta I_{C} \approx 0.03$, 
$\delta A_{I_{C}} \approx 0.42$, $\delta BC_{I_{C}} \approx 0.01$ and $\delta (5 \log D) \approx 0.08$.  
The effect of distance uncertainty on the relative positions of the stars 
was estimated with the assumption that the scatter of the distances of 
stars is same as their largest projected  separation, i.e. $\sim$\,7\,pc.

The luminosity of \textit{IRAS}~02086+7600 over the wavelength interval  
0.55--135\,$\mu$m was calculated by integrating the available dereddened 
fluxes. Applying a long-wavelength correction as proposed by \citet{Kenyon90}
for objects with SED peaks near $100\,\mu$m, $L_{>135} \approx 0.86\,L_{100}$, 
and neglecting the luminosity below 0.55\,$\mu$m, we obtained  a total luminosity of
$L_{tot} = 1.04\,L_{\sun}$. In addition to the bolometric luminosity of the central star
($L^{*}_{bol}$), this total luminosity includes the luminosity of the accretion shock 
($L_{acc,shock}$), and those generated and reprocessed in the 
disc ($L_{acc,disc}$ and $L_{rep,disc}$). In order to obtain $L^{*}_{bol}$, 
contributions of $L_{acc,shock}$, $L_{acc,disc}$, and $L_{rep,disc}$ 
have to be subtracted from $L_{tot}$. To this end we utilized two empirical 
relationships, resulted from comprehensive studies of large YSO samples.
First, it was shown by \citet{WH04}, that the total luminosity of Class~I sources
can be approximated as $L_{tot} = 1.08\,L^{*}_{bol} + 1.58\,L_{acc,shock}$, where
$L_{acc,shock} = (G M_{*} \dot{M}) / R_{*}$. The second relationship is that 
established by \citet{Muzerolle} between $\dot{M}$ and the luminosity of the 
\mbox{Ca\,{\sc ii}}\,$\lambda$\,8542 line. Following their method, 
the measured $EW(\mbox{Ca\,{\sc ii}}\,\lambda\,8542) = 10$\,\AA \ resulted in
$\dot{M} \approx 5.7 \times 10^{-9} M_{\sun} / yr$, 
comparable to the median value $\dot{M} = 7.9\times 10^{-9} M_{\sun} / yr$, 
obtained by \citet{WH04} for the Class~I objects of Taurus. With the assumptions  
$M_{*} = 0.8\,M_{\sun}$, and $R_{*} = 2\,R_{\sun}$ the obtained mass 
accretion rate led to $L_{acc,shock} = 0.07\,L_{\sun}$ and  $L^{*}_{bol} = 0.86\,L_{\sun}$.
The resulting  temperature and luminosity of \textit{IRAS}~02086+7600 is 
plotted as the open circle in Fig.~\ref{Fig_hrd}. 
The  uncertainty was calculated as in the case of OKS\,H$\alpha$ stars. 

Evolutionary tracks and isochrones, as well as the position of the birthline 
and zero-age main-sequence \citep{PS99} are also shown in Fig.~\ref{Fig_hrd}. 
We obtained masses 0.8\,M$_{\sun}$ and 0.2\,M$_{\sun}$ for OKS\,H$\alpha$\,6\,N 
and OKS\,H$\alpha$\,6\,S, respectively. Our data suggest that
within the accuracy  both components are coeval, $\sim$\,3--5~million years old.  
OKS\,H$\alpha$\,5, OKS\,H$\alpha$\,16, and \textit{IRAS}~02086+7600 lie 
close to the $10^{6}$-yr isochrone. 
The weak-line T~Tauri stars identified by \citet{TNKF05}
near L\,1333 are also plotted in Fig.~\ref{Fig_hrd} (see Sect.~\ref{Sect_3.4}). 
Table~\ref{Tab_result} summarizes our main results:
$A_{V}$, $T_\rmn{eff}$, $L / L_{\sun}$, as well as the masses in
$M_{\sun}$ and ages in $10^{6}$\,yrs, read from Fig.~\ref{Fig_hrd}. 

\begin{table}
\begin{minipage}{80mm}
\caption{Properties of the pre-main-sequence stars associated with 
L\,1333, derived from spectroscopic and photometric data}
\label{Tab_result}
{\footnotesize
\begin{tabular}{lccccc}
\hline
Star & $A_\rmn{V}$ & $T_\rmn{eff}$ & $L$ & Mass & Age  \\
 & (mag) &  (K) & ($L_{\sun}$) &  ($M_{\sun}$) & (10${^6}$ yr) \\
\hline
\textit{IRAS}~02086 & 3.76\,(1.2:) & 4000 & 0.86 &  0.8~ &  1.5 \\
OKS\,H$\alpha$\,5  &  2.57\,(0.42) &  3570 & 0.58 &  0.3~ & 1.0 \\
OKS\,H$\alpha$\,6\,N & 0.71\,(0.42) &  4060 &  0.49 &  0.8~ & 5.0  \\
OKS\,H$\alpha$\,6\,S & 1.29\,(0.55) &  3410 &  0.11 &  0.15 & 3.0  \\
OKS\,H$\alpha$\,16  & 5.28\,(0.50) &  3870 &   0.93 &  0.4~ & 1.0 \\
\hline
\end{tabular}}
\end{minipage}
\end{table}

\subsection{Large-scale environment of L\,1333}
\label{Sect_3.3}

\begin{figure*}
\centering{
\includegraphics[width=14cm]{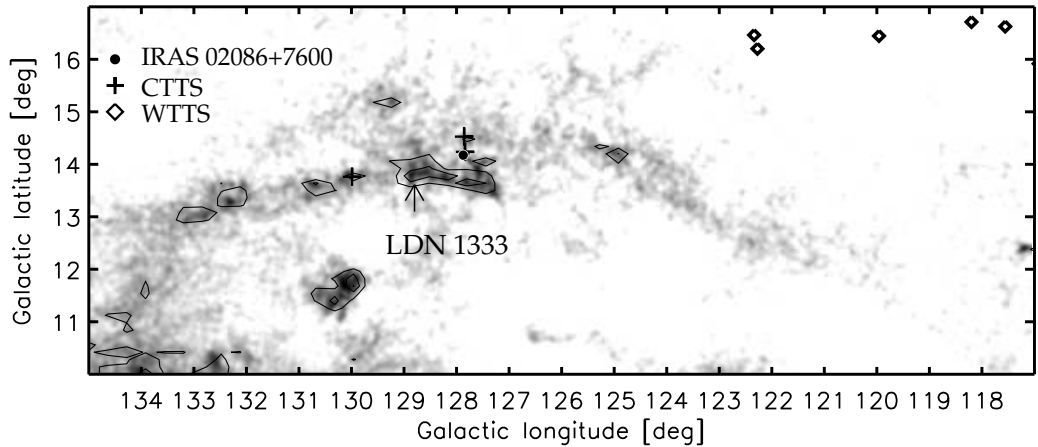}}
\caption{Large-scale distribution of the visual extinction around L\,1333, adopted from
\citet{DUK}. Contours are drawn at $A_V=1.0~mag$ and $1.5~mag$. 
YSOs associated with the  L\,1333 molecular complex, and the  weak-line T~Tauri stars
identified by \citet{TNKF05} are indicated.}
\label{Fig_ext}
\end{figure*}

The map of the visual  extinction of the region $117\degr \lid l \lid 135\degr$ and
$+10\degr \lid b \lid +17\degr$, taken from the
{\sl Atlas and Catalog of Dark Clouds\/} by \citet{DUK} and displayed
in Fig.~\ref{Fig_ext}, shows that L\,1333 is near the middle of a long, diffuse 
filamentary cloud complex spanning from $l \sim 120\degr$ to $l\sim134\degr$,  
far beyond the limits of the molecular observations presented in Paper~I. 
The dark cloud seen at $127\degr \la l \la 131\degr$ is catalogued as DUK~853 by \citet{DUK} 
and contains eight clumps ({\sl P1--P8\/} in the order of decreasing mass). L\,1333 as catalogued by 
\citet{Lynds}  corresponds to the largest clump~{\sl P1\/}. 
\textit{IRAS}~02086+7600 and OKS\,H$\alpha$\,5 are located at the high-latitude 
edge of the second largest clump~{\sl P2\/}, and OKS\,H$\alpha$\,6 is projected on 
the edge of the small clump {\sl P7\/} at the highest latitude side of DUK~853. 
OKS\,H$\alpha$\,16 is projected near the centre of
clump~{\sl P4\/} of the same dark cloud.
\citeauthor{DUK}'s catalogue provides an opportunity to derive the masses  
of the clouds and their clumps. The total mass of the 
clouds within the diffuse filament between $l \sim 120\degr$ and $l \sim 134\degr$,  
derived from the visual extinction, is $\sim$\,2300\,M$_{\odot}$. 
Clump masses range between 2 and 30\,M$_{\odot}$. 

In order to assess the star forming history of the whole region we also plotted in Fig.~\ref{Fig_ext}
the weak-line T~Tauri stars identified by \citet{TNKF05}, and lying 
far from any dark cloud. \citeauthor{TNKF05} suggest that the parent clouds of these 
stars might have been connected to the  L\,1333 complex. In order to properly compare 
the ages of these WTTSs with those of our CTTSs, we plotted their data, taken from 
\citeauthor{TNKF05}'s Table~3, in Fig.~\ref{Fig_hrd}
(\citeauthor{TNKF05} used isochrones of \citet{DM94}, giving somewhat different results.).
The ages of the WTTSs, assuming a distance of 200\,pc, are between 3 and 10
million years, with the youngest one at the highest longitude end of the chain, 
and the oldest on the low-latitude end. 

\subsection{A possible scenario of star formation}
\label{Sect_3.4}

Comparison of the properties of dense C$^{18}$O cores of L\,1333 with those of
other nearby star forming clouds have shown these cores to be smaller and less
massive than the similar regions of Taurus, Ophiuchus, Lupus and Chamaeleon clouds
\citep[Paper~I;][]{MHY,TOMF02}. 
Star formation in such an environment is thought to be assisted by some external trigger, 
and the filamentary clouds themselves have probably been created by large-scale motions 
of the interstellar gas. The most plausible scenario, suggested by the arc-like structure
is, that energetic stellar winds and/or supernova explosions of high-mass stars at 
lower galactic latitudes lifted the gas above the galactic plane and compressed 
it to form stars. In this case the apparent filament is a projection of a shell,
and its line-of-sight extent may be comparable to its length. 
The distribution of YSOs relative to the clouds does not support this scenario.
The young stars of L\,1333 are located at the high-latitude side of the cloud, with the 
oldest member, OKS\,H$\alpha$\,6, lying farthest from the cloud. The lack of
H$\alpha$ emission stars, as well as YSO-like \textit{IRAS} and 2MASS point sources 
on the low-latitude side of the filament suggests star formation propagating 
toward lower latitudes, and a source of trigger at higher galactic latitudes.

A possible candidate trigger source
is the collision of high velocity gas with the giant radio continuum emitting region 
{\sl Loop~III\/}, described by  \citet{Verschuur}. 
Our target objects are located  near the far side of
Loop~III \citep{Berkhuijsen,Spoelstra}. \citet{Verschuur} has shown that
Loop~III collided with high velocity gas originating from a galactic supershell some 
$7 \times 10^{5}$ ago.
The collision has been well modelled for latitudes $b > 20\degr$. At lower latitudes, however, 
the behaviour of the supershell  and its collision with the local interstellar matter
has not yet been studied. In order to reveal the geometry of the possible collision
in the latitude range $10\degr$--$20\degr$, the velocity distributions
of both molecular and atomic gas have to be studied in detail. Closer to the 
galactic plane the high velocity gas of the supershell might have 
decelerated before reaching our galactic neighbourhood. 

In this scenario the high-density regions, created by the colliding surfaces, 
have small line-of-sight extent.
The ages obtained for \textit{IRAS}~02086+7600, OKS\,H$\alpha$\,5, and 
OKS\,H$\alpha$\,16, taking into account their accuracies, 
support this scenario. OKS\,H$\alpha$\,6 was, however, born apparently before the collision. 
The weak-line T~Tauri stars  to the west of the cloud complex make the 
pattern of star formation of this region even more complicated. They indicate 
a prolonged star formation in the region. 
The age distribution of the WTTSs suggests star formation propagating from lower 
to higher galactic longitudes. More accurate age determinations and more detailed 
mapping of molecular velocity distribution are needed to clarify the picture.

\section{Conclusions}
\label{Sect_4}

We identified five low-mass YSOs in the small filamentary molecular complex 
associated with the dark cloud Lynds~1333. Their masses are in the interval 
0.15--0.8\,M$_{\sun}$, and they are 1--5 million years old. We confirmed that 
\textit{IRAS}~02086+7600 is a Class~I YSO associated with the L\,1333 complex,
and found its age to be comparable to those of the CTTSs born in the same cloud.
The relative distribution of YSOs and clouds suggests that the star formation 
might have been triggered by the collision of high velocity gas with Loop~III.

\section*{Acknowledgements}

This work is partly based on observations with Nordic Optical Telescope  
operated on the island of La Palma jointly by Denmark, Finland, 
Iceland, Norway, and Sweden, in the Spanish Observatorio 
del Roque de los Muchachos of the Instituto de Astrof\'{i}sica de Canarias.
The data presented here have been taken using ALFOSC, which is owned
by the Instituto de Astrof\'{i}sica de Andaluc{\'i}a (IAA) and operated at the
Nordic Optical Telescope under agreement between IAA and the NBIfAFG of 
the Astronomical Observatory of Copenhagen. Our results are partly based on 
observations obtained at the Centro Astron\'omico Hispano Alem\'an (CAHA) at 
Calar Alto, operated jointly by the Max-Planck Institut f\"ur Astronomie and the 
Instituto de Astrof\'{\i}sica de Andaluc\'{\i}a (CSIC).
We are indebted to Francesco Palla for sending his data set on pre-main 
sequence evolution, and to L\'aszl\'o Sza\-bados for careful reading of the
manuscript. Financial support from the Hungarian OTKA grants T034584, T037508, 
TS049872, T042509, and T049082 is acknowledged. SN acknowledges 
support from the Chilean \emph{Centro de Astrof\'isica} FONDAP No.~15010003 and
Serbian Ministry of Science and Environmental Protection grant No.~146016.

\end{document}